\documentclass[twocolumn]{aastex63}

\newcommand{\lya}{Ly$\alpha$}
\newcommand{\oiii}{[O\,{\sc iii}]}
\newcommand{\ha}{H$\alpha$}
\newcommand{\hb}{H$\beta$}
\newcommand{\hst}{{\it HST}}

\received{2019 Nov 20}
\revised{2019 Dec 10}
\accepted{2019 Dec 21}

\shorttitle{Very Blue Galaxies at $z\simeq6$}
\shortauthors{Jiang et al.}

\begin{document}

\title{Luminous \lya\ Emitters with Very Blue UV-Continuum Slopes at Redshift $5.7\le z \le 6.6$}


\author[0000-0003-4176-6486]{Linhua Jiang}
\affiliation{Kavli Institute for Astronomy and Astrophysics, Peking University, Beijing 100871, China; jiangKIAA@pku.edu.cn}

\author{Seth H. Cohen}
\affiliation{School of Earth and Space Exploration, Arizona State University, Tempe, AZ 85287-1504, USA}

\author{Rogier A. Windhorst}
\affiliation{School of Earth and Space Exploration, Arizona State University, Tempe, AZ 85287-1504, USA}

\author{Eiichi Egami}
\affiliation{Steward Observatory, University of Arizona, 933 North Cherry Avenue, Tucson, AZ 85721, USA}

\author{Kristian Finlator}
\affiliation{New Mexico State University, Las Cruces, NM 88003, USA}

\author{Daniel Schaerer}
\affiliation{Geneva Observatory, University of Geneva, Ch. des Maillettes 51, 1290 Versoix, Switzerland}

\author{Fengwu Sun}
\affiliation{Steward Observatory, University of Arizona, 933 North Cherry Avenue, Tucson, AZ 85721, USA}


\begin{abstract}

We study six luminous \lya\ emitters (LAEs) with very blue rest-frame UV continua at $5.7\le z \le 6.6$. These LAEs have previous \hst\ and {\it Spitzer} IRAC observations. Combining our newly acquired \hst\ images, we find that their UV-continuum slopes $\beta$ are in a range of $-3.4\le \beta \le -2.6$. Unlike previous, tentative detections of $\beta \simeq -3$ in photometrically selected, low-luminosity ($M_{\rm UV} \ge -18.5$ mag) galaxies, our LAEs are spectroscopically confirmed and luminous ($-22<M_{\rm UV}<-20$ mag). We model their broadband spectral energy distributions (SEDs), and find that two $\beta\simeq-2.6\pm0.2$ galaxies can be well fitted with young and dust-free stellar populations. However, it becomes increasingly difficult to fit bluer galaxies. We explore further interpretations by including non-zero LyC escape fraction $f_{\rm esc}$, very low metallicities, and/or AGN contributions. Assuming $f_{\rm esc}\simeq0.2$, we achieve the bluest slopes $\beta\simeq-2.7$ when nebular emission is considered. This can nearly explain the SEDs of two galaxies with $\beta\simeq-2.8$ and --2.9 ($\sigma_{\beta}=0.15$). Larger $f_{\rm esc}$ values and very low metallicities are not favored by the strong nebular line emission (evidenced by the IRAC flux) or the observed (IRAC~1 -- IRAC~2) color. Finally, we find that the $\beta\simeq-2.9$ galaxy can potentially be well explained by the combination of a very young population with a high $f_{\rm esc}$ ($\ge0.5$) and an old, dusty population. We are not able to produce two $\beta \simeq -3.4 \pm0.4$ galaxies. Future deep spectroscopic observations are needed to fully understand these galaxies.

\end{abstract}


\section{Introduction} \label{sec:intro}

In recent years, a number of high-redshift galaxies have been discovered, thanks to the advances of instrumentation on the {\it Hubble Space Telescope} (\hst) and large ground-based telescopes. These galaxies have played an important role in studies of galaxy formation and evolution at early epochs. Most known galaxies at $z\ge6$ are photometrically selected Lyman-break galaxies (LBGs) or candidates using the dropout technique \citep[e.g.,][]{mcl11,yan12,ellis13,bou15,ono18}. Only a small fraction of them have been spectroscopically confirmed \citep[e.g.,][]{fin13,oesch15,jiang17,lap17,stark17}. The narrowband (or \lya) technique provides a complementary method to find high-redshift galaxies, and narrowband observations were mostly done by large-area ground-based observations \citep[e.g.,][]{kas11,ota17,zheng17,ouchi18,hu19}. While ground-based observations provide samples of luminous galaxies, the majority of faint galaxies have to come from deep \hst\ observations. 

The physical properties of high-redshift galaxies are also being investigated. The combination of the \hst\ and {\it Spitzer} (IRAC~1 and IRAC~2) observations covers the rest-frame UV and optical ranges for these galaxies, and is thus critical to characterize these objects and measure their stellar populations. The UV continuum slope $\beta$ ($f_{\lambda} \propto {\lambda}^{\beta}$) is of particular interest. It provides key information to constrain young stellar populations, yet it does not require {\it Spitzer} observations. Previous \hst\ observations have tentatively found extremely steep slopes $\beta\simeq-3$ in photometrically selected, very faint galaxies at $z \ge 6$ \citep[e.g.,][]{bou10,lab10}. Such slopes have not been reported in lower-redshift galaxies, nor have they been predicted by cosmological simulations \citep[e.g.,][]{finlator11,cev19}. On the other hand, this finding is controversial, as others have demonstrated the likelihood that the extreme slopes could be caused by contamination and bias due to the nature of photometrically selected samples \citep[e.g.,][]{mcl11,dun12,dun13}. As the debate goes on, the key is to construct a reliable sample of galaxies with $\beta\simeq-3$.

We have carried out a detailed study of a sample of 67 spectroscopically confirmed galaxies at $z\simeq6$ \citep{jiang13a,jiang13b,jiang16}, based on our Subaru optical, \hst\ near-IR, and {\it Spitzer} mid-IR images. These galaxies have blue UV continua with a median slope $\beta \simeq -2.3$ at $M_{\rm UV} < -19.5$ mag. In particular, a non-negligible fraction of them exhibit extreme slopes $\beta \simeq -3$ (see Figure 3 in Jiang et al. 2013a). Due to their brightness and secure redshifts, the $\beta$ measurements were not subject to the contamination or bias mentioned above. The uncertainties $\sigma_{\beta}$ are around $0.3-0.5$, mainly due to the short UV-continuum baseline or large photometric errors. In order to reduce $\sigma_{\beta}$, we have obtained new \hst\ WFC3 data from \hst\ Cycle 25 for seven relatively bright and promising candidates of extremely blue galaxies selected from the above sample.

In this paper, we present the new \hst\ observations and confirm very steep slopes in six of the seven galaxies. The layout of the paper is as follows. In Section \ref{sec:data}, we introduce our \hst\ observations and data reduction. In Section \ref{sec:results}, we present our main results of the $\beta$ measurements. In Section \ref{sec:disc}, we model the spectral energy distributions (SEDs) these galaxies and discuss our results. We summarize our paper in Section \ref{sec:sum}. Throughout the paper, all magnitudes are expressed on the AB system. The \hst\ F105W, F110W, F125W, F140W, and F160W bands are denoted as $Y_{105}$, $J_{110}$, $J_{125}$, $H_{140}$, and $H_{160}$, respectively. We use a $\Lambda$-dominated flat cosmology with $H_0=68$ km s$^{-1}$ Mpc$^{-1}$, $\Omega_{m}=0.3$, and $\Omega_{\Lambda}=0.7$.

\section{Observations and data reduction} \label{sec:data}

\begin{deluxetable*}{cccccccccc}
\tablecaption{The $z\simeq6$ galaxy sample}
\tablewidth{0pt}
\tablehead{
\colhead{ID} & \colhead{Redshift} & \colhead{Slope} & \colhead{$z'$} & \colhead{$Y_{105}$} & \colhead{$J_{125}$} & \colhead{$H_{160}$}  & \colhead{$K_{\rm s}$} & \colhead{IRAC 1} & \colhead{IRAC 2} \\
\nocolhead{} & \nocolhead{} & \colhead{($\beta$)} & \colhead{(mag)} & \colhead{(mag)} & \colhead{(mag)} & \colhead{(mag)} & \colhead{(mag)} & \colhead{(mag)} & \colhead{(mag)} }
\colnumbers
\startdata
ID07 & 5.691 & --3.38$\pm$0.35 & 26.24$\pm$0.14 & 26.66$\pm$0.14 & 26.74$\pm$0.12 & 27.10$\pm$0.17 &       $\ldots$        &        $\ldots$       & $\ldots$ \\
ID28 & 6.042 & --2.57$\pm$0.20 & 25.65$\pm$0.06 & 26.12$\pm$0.08 & 26.30$\pm$0.11 & 26.40$\pm$0.10 &       $\ldots$        & 25.59$\pm$0.39 & $\ldots$ \\
ID30 & 6.062 & --2.60$\pm$0.25 & 25.72$\pm$0.07 & 25.68$\pm$0.09 & 25.94$\pm$0.11 & 26.04$\pm$0.13 &       $\ldots$        & 25.50$\pm$0.35 & $\ldots$ \\
ID43 & 6.542 & --3.39$\pm$0.41 &       $\ldots$        & 25.94$\pm$0.13 & 26.29$\pm$0.07 & 26.55$\pm$0.11 &       $\ldots$        & 25.30$\pm$0.24 & $\ldots$ \\
ID61 & 6.599 & --1.41$\pm$0.30 &       $\ldots$        & 26.02$\pm$0.09 & 26.01$\pm$0.12 & 25.90$\pm$0.07 &       $\ldots$        & 24.08$\pm$0.07 & 24.50$\pm$0.18 \\
ID63 & 6.027 & --2.89$\pm$0.16 & 24.73$\pm$0.06 & 25.32$\pm$0.15 & 25.19$\pm$0.05 & 25.42$\pm$0.08 & $25.72\pm0.17$ & 24.49$\pm$0.12 & 24.97$\pm$0.25 \\
ID64 & 6.120 & --2.77$\pm$0.15 & 25.10$\pm$0.06 & 25.43$\pm$0.13 & 25.36$\pm$0.06 & 25.69$\pm$0.07 & $25.76\pm0.17$ & 24.90$\pm$0.11 & $\ldots$ \\
\enddata
\tablecomments{The ID numbers in Column 1 correspond to the numbers in Table 1 in \citet{jiang13a}. For ID28, the photometry in Column 5 is for $J_{110}$, and the photometry in Column 6 is for $H_{140}$. The $K_{\rm s}$-band photometry in Column 8 is from \citet{gal13}. The IRAC photometry in Columns 9 and 10 is from \citet{jiang16}.}
\end{deluxetable*}

We first briefly review our previous \hst\ observations of 67 galaxies at $5.7\le z \le 6.6$ \citep{jiang13a}. These galaxies were spectroscopically confirmed via \lya\ emission lines. They represent the most luminous $z\simeq6$ galaxies in terms of \lya\ emission or UV-continuum emission. Most of them are located in the Subaru Deep Field \citep[SDF;][]{kas04}, and the remaining are located in the Subaru XMM-Newton Deep Survey \citep[SXDS;][]{fur08} field. They have deep optical images from the Subaru telescope. They also have deep near-IR and mid-IR images from our previous \hst\ and {\it Spitzer} programs. The \hst\ observations of the SDF galaxies were done with a mix of strategies. Most galaxies were observed in the WFC3 $J_{125}$ and $H_{160}$ bands, with a depth of two orbits per band. For the galaxies in the SXDS field, we have used the CANDELS \citep{gro11,koe11} $J_{125}$- and $H_{160}$-band images. Based on these data, we previously found that our galaxies have blue UV slopes with a median value of $\beta \simeq -2.3$. In addition, a small fraction of them show extreme slopes of $\beta \simeq -3$, but the errors were relatively large ($\sigma_{\beta}=0.3\sim0.5$).

In order to study these extreme galaxies, we have carried out new \hst\ WFC3 observations of seven LAEs (Table 1) in \hst\ Cycle 25 program 15137 (PI: L. Jiang). These were selected from the above sample to be relatively bright ($>10\sigma$ detection in $J_{125}$) with $\beta \le -2.8$. We took advantage of the existing data and improved the $\beta$ measurements by extending the UV continuum baseline, increasing the spectral sampling, and/or improving the photometry in $H_{160}$. We observed all objects except ID28 in $Y_{105}$, which provides a longer UV continuum baseline. The depth was one \hst\ orbit per object. Because of the very blue slopes, the galaxies were expected to be brighter in $Y_{105}$, but much fainter in $H_{160}$. Therefore, we obtained new $H_{160}$-band images for four objects (ID07, ID28, ID30, and ID61), and the depth was also one orbit per object. ID28 already had images in $J_{110}$ and $H_{160}$, so we obtained one-orbit $H_{140}$-band images to improve the spectral sampling.

We used the software package DrizzlePac and followed the standard procedure to reduce our \hst\ images. For each galaxy, all  images were drizzled onto the same WCS system. The new $Y_{105}$-band images of the two SXDS galaxies were drizzled to match the WCS in the CANDELS images. The `pixfrac' parameter in DrizzlePac was set to be 0.8 and the output plate scale is $0\farcs06$ pixel$^{-1}$. We combined the old and new images in the individual bands. The final depth is roughly one orbit in $Y_{105}$, two orbits in $J_{125}$, and three orbits in $H_{160}$, corresponding to the $10\sigma$ detection limits (within a $0\farcs6$ or 10-pixel circular aperture in diameter) of $\sim26.3$, 26.7, and 26.6 mag, respectively.

In addition to the \hst\ images, we have used deep $z'$-band imaging data from Subaru Suprime-Cam to extend the UV-continuum baseline when calculating $\beta$. These data have been used to select our galaxies, and the details were described in \citet{jiang13a}. We have slightly improved the $z'$-band photometry of the galaxies. Furthermore, we have also used deep ground-based $K_{\rm s}$-band data to further extend the UV baseline for two SXDS galaxies (see Section \ref{sec:results}).

\section{Results} \label{sec:results}

\begin{figure}[t]
\plotone{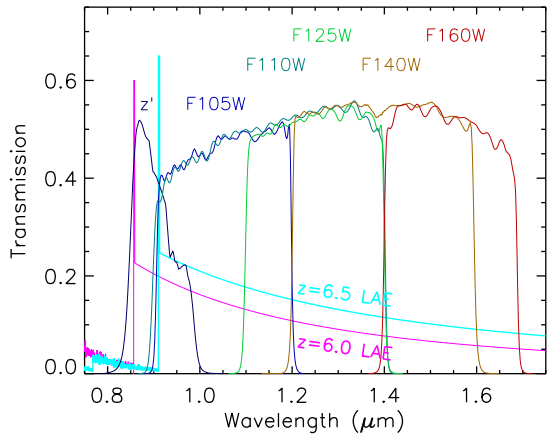}
\caption{Transmission curves of the filters that have been used to measure the UV-continuum slopes of the galaxies. The magenta and cyan spectra represent two model LAEs at $z=6.0$ and 6.5, respectively. The \lya\ emission line enters the $z'$ band for $z\simeq6$ galaxies, and enters the $Y_{105}$ and $J_{110}$ bands for $z\simeq6.5$ galaxies.
\label{fig:filters}}
\end{figure}

In this section, we will measure the photometry of the galaxies in multiple bands, and then calculate their UV-continuum slopes using the multi-band photometric data. We will show that our slopes are not estimated from a single UV color as many previous studies did for high-redshift galaxies. Instead, they are measured from 3--4 bands (Figures \ref{fig:filters} and \ref{fig:calBeta}). 

\subsection{Photometry} \label{subsec:results1}

We run SExtractor \citep{ber96} to measure the photometry. The procedure is similar to those in the literature. Usually \hst\ images in a longer wavelength band have a larger size (or FWHM) of the point-spread function (PSF). But this is not the case in our images. We performed sub-pixel dithering during the \hst\ observations, which can improve the PSF. As mentioned earlier, the typical numbers of the \hst\ orbits for our galaxies are 1, 2, and 3 in $Y_{105}$, $J_{125}$, and $H_{160}$, respectively. This results in nearly consistent PSF sizes (within the errors) in the different bands \citep[see e.g.,][]{win11}. For example, the PSF FWHM values in the three bands, measured from bright stars by SExtractor, are $0\farcs22\pm0\farcs01$, $0\farcs21\pm0\farcs02$, and $0\farcs22\pm0\farcs02$, respectively. Therefore, we skip PSF matching here. To do photometry for a given object, we first stack (inverse-variance weighted average) all its \hst\ images to make a combined image. We then run SExtractor in the dual image mode, using the combined image as the `detection' image. We calculate the total magnitude within a MAG\_AUTO elliptical aperture with PHOT\_AUTOPARAMS values of 1.8 and 2.5. The aperture corrections are computed using standard (larger) PHOT\_AUTOPARAMS values of 2.5 and 3.5. The results are shown in Table 1.

The photometric uncertainties are also measured by SExtractor. We have converted the weight or variance maps (obtained from DrizzlePac) to rms images following the procedure of \citet{dic04} that includes correlated noise. The rms images are then fed to SExtractor. The average background noise is estimated using a $2\farcs4$-wide annulus (larger than the default value by 67\%), and the total errors are calculated within the adopted aperture.  We evaluate our error measurements using mock data. For a given galaxy in a given image, we generate 100 objects with the same brightness and shape as the galaxy, and put them randomly on the image. We then do photometry using SExtractor as we did for real objects, and compute the mean and standard deviation of the measurements. The new results are consistent with the results above.

The aperture size that we have adopted is roughly $0\farcs6$ in diameter, larger than those in the majority of the previous studies of $z \ge 6$ galaxies. This results in relatively larger measurement errors. On the other hand, this conservative approach reduces the effect from any possible PSF mismatches between different bands. Our galaxies are isolated in the \hst\ images, and thus no obvious neighbors are included in the photometry. The two galaxies ID63 and ID64 are bright enough to test if the magnitudes that we have measured are close to `real' total magnitudes. We measure their photometry again within a very large aperture of $1\arcsec$, and the results are well consistent with the values given in Table 1. 

The $z'$-band photometry is done within a $2\arcsec$ circular aperture in diameter using SExtractor. Aperture corrections are derived from bright stars. All galaxies except ID61 are compact in the \hst\ images, so they are point-like sources in the ground-based $z'$-band images. Therefore, the usage of a circular aperture and the aperture corrections from stars are robust. The results are shown in Column 6 of Table 1. ID61 is likely an interacting system \citep[see also][]{jiang13a}, and its UV slope is not blue (Table 1).

\subsection{UV continuum slopes} \label{subsec:results2}

\begin{figure}[t]
\plotone{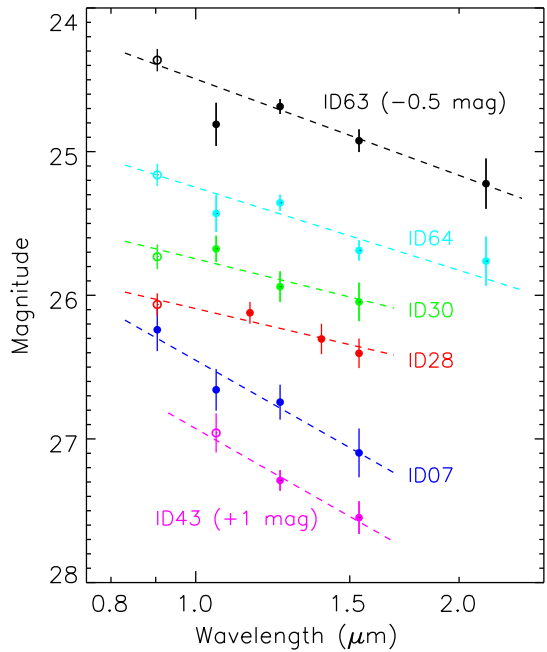}
\caption{Measurement of the UV-continuum slopes for 6 galaxies with $\beta\le-2.6$. The color-coded symbols with $1\sigma$ error bars represent the photometric data points. The data points for ID63 and ID43 have been shifted for clarity. The first data points (open circles) have been corrected for the IGM absorption and \lya\ emission. The dashed lines are the best fits to the data in 3--4 bands. 
\label{fig:calBeta}}
\end{figure}

We measure UV-continuum slopes by fitting a power-law ($f_{\lambda} \propto {\lambda}^{\beta}$) to the photometric data above. The procedure is the same as what we did in \citet{jiang13a}. Because AB magnitude $m_{\rm AB}$ and $\rm log(\lambda)$ follow the linear relation $m_{\rm AB} \propto (\beta+2) \times {\rm log}(\lambda)$, we actually perform a linear fit on $m_{\rm AB}$ and $\rm log(\lambda)$ with the $m_{\rm AB}$ errors included, i.e., we do a standard (weighted) least-square linear regression by minimizing the $\chi^2$ error statistic. The errors propagate to the final parameters. The results are listed in Table 1. They are also illustrated in Figure \ref{fig:calBeta}. All galaxies except ID61 have steep slopes with $\beta\le-2.6$.

The \lya\ emission line enters the $z'$ band for $z\simeq6$ galaxies, and enters the $Y_{105}$ and $J_{110}$ bands for $z\simeq6.5$ galaxies (see Figure \ref{fig:filters}). Hence, the \lya\ emission increases the relevant broadband flux. The \lya\ line flux of our galaxies was provided in \citet{jiang13a}. Note that we do not use the $z'$ band for $z\simeq6.5$ galaxies. On the other hand, the flux blueward of \lya\ is nearly completely absorbed by the IGM at such high redshifts, which decreases the relevant broadband flux. These effects have been taken into account in this paper. The data points shown in Figure \ref{fig:calBeta} have also been corrected for the IGM absorption and \lya\ emission. The details are as follows. To correct for the IGM absorption, we use the model in \citet{mcg13}. Figure \ref{fig:filters} shows two LAE model spectra with the IGM absorption at $z=6.0$ and 6.5. Since the \lya\ emission line of a $z \approx 6.0$ (or 6.5) galaxy is at the edge of the $z'$ (or $Y_{105}$) filter, the uncertainty of the correction for the IGM absorption is negligible.

We use our spectroscopic redshifts and reliable \lya\ flux measurements to remove the \lya\ contribution in the $z'$- or  $Y_{105}$-band photometry before we calculate UV continuum slopes or perform SED modeling. The system transmission, including the filter and detector, is considered. The typical rest-frame equivalent width of \lya\ in our sample is about 50 \AA. The \lya\ contribution to the $z'$-band photometry is roughly 20\% (10\% in $Y_{105}$), because the two filters are both very wide. Therefore, a 20\% uncertainty on the \lya\ line flux measurement causes a 4\% uncertainty in the $z'$-band photometry, or a 2\% uncertainty in $Y_{105}$. We have added an additional error of 0.05 mag in quadrature to the relevant bands to include extra uncertainties introduced in this step.

The two SXDS galaxies ID63 and ID64 are the brightest in our sample, and can be well detected in the $K$ band from the ground. We have adopted the VLT HAWKI-I $K_{\rm s}$-band photometry from \citet{gal13}. The total magnitudes are $25.72\pm0.17$ and $25.76\pm0.17$ mag, respectively. We also find that the $Y_{105}$-band data points of the two objects largely deviate from the linear fits. The two galaxies are bright, compact, and isolated in the \hst\ images, so their photometric measurements are robust. We check the four individual exposures (in four dither positions) for each galaxy, and find that the flux varies by $\sim$0.5 mag. Due to this unusually large variation, the actual errors (0.13-0.15 mag shown in Table 1) are much larger than the measurement errors (0.05 mag). 

ID61 is the only galaxy that obviously has an elongated shape. It seems to be an interacting system, and its minor component is brighter at longer wavelengths. Our current \hst\ images do not have enough resolution to reliably separate the two components. The magnitudes quoted in Table 1 are the total magnitudes, and the slope ($\beta\simeq-1.41$) is not  steep. We will not discuss this galaxy later.

Among the six blue galaxies, two galaxies ID28 and ID30 have $\beta\approx-2.6\pm0.2$. Their slopes become flatter compared to our previous measurements ($\beta \approx -2.8$ and --3.5 with $\sigma_{\beta} \approx -0.4$) in \citet{jiang13a}. This is mainly due to the significant improvement of the photometry in $H_{160}$. The new photometry shows that the two objects become much `brighter' in $H_{160}$, resulting in the flatter slopes. If we do not use their $z'$-band data, the slopes become $\beta\approx-2.9\pm0.3$. Two galaxies ID63 and ID64 respectively have $\beta\approx-2.9$ and --2.8 with $\sigma_\beta\approx0.15$, which are consistent with our previous measurements $\beta\simeq-2.8$ and --3.0, but with reduced errors. The main reason for the consistent results is that the two objects are bright, and the previous measurements were reliable. If we do not use their $z'$-band data, the slopes become $\beta\approx-2.8\pm0.2$. And if we further exclude the $Y_{105}$-band data (these data are poor as mentioned above), the slopes become $\beta\approx-3.0\pm0.3$. All these measurements are consistent within $1\sigma$.

The other two galaxies ID07 and ID43 have very steep slopes $\beta \approx -3.4$. But the errors $\sigma_{\beta} \approx 0.4$ are also larger, because they are relatively faint. For ID43, if we exclude its $Y_{105}$-band data, the slope becomes $\beta\approx-3.2\pm0.6$. It would be important to improve their photometry or obtain their spectra in future observations.

\begin{figure}[t]
\plotone{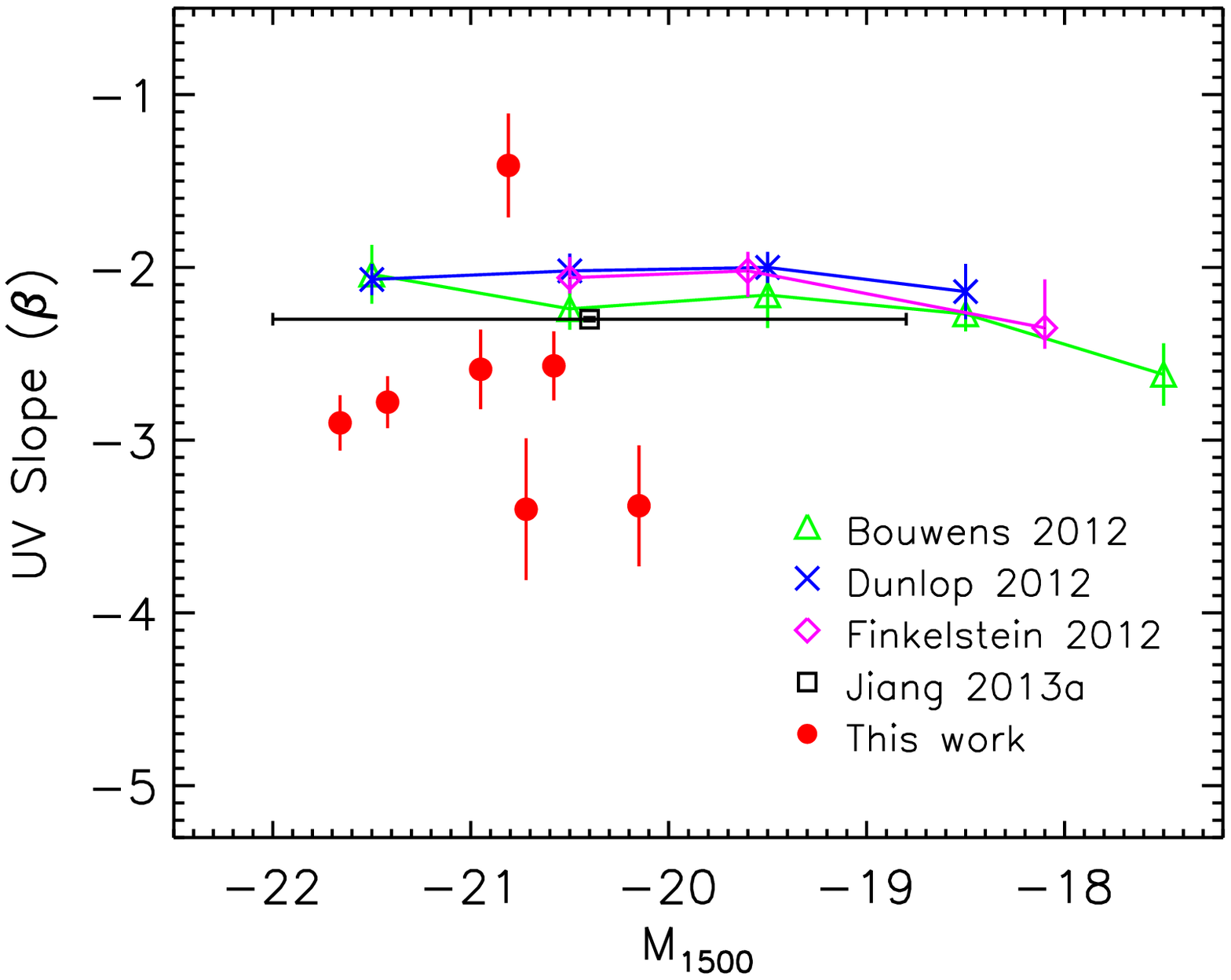}
\caption{Comparison of $\beta$ with previous studies. The green, blue, magenta symbols represent previous $\beta$ measurements from \hst\ \citep{bou12,dun12,fin12}. They are mean values in magnitude bins for photometrically selected $z\simeq6$ LBGs . The black square represents the average slope $\beta=-2.3$ in the \citet{jiang13a} sample, and its horizontal error bar indicates the rough luminosity range covered by the sample. Our LAEs are relatively luminous, and six of them are bluer than the average slope in the same luminosity range. 
\label{fig:comBeta}}
\end{figure}

Figure \ref{fig:comBeta} compares our results with previous measurements of $\beta$ in photometrically selected $z\simeq6$ LBGs from \hst. Our six galaxies are luminous (by selection) with $M_{1500}$ between --22 and --20 mag. In this luminosity range, typical star-forming galaxies have slopes around $\beta \simeq -2$ or even redder \citep{jiang13a,mat19}. Our previous results show that LAEs have an average slope of $\beta\simeq-2.3$ (black square in Figure \ref{fig:comBeta}) in the range of $-22<M_{1500}<-19$ mag. In addition, a small fraction of LAEs have $\beta \simeq -3$, but with larger errors. Such extreme slopes were only reported in very faint ($J/H\simeq28$ mag or $M_{1500}\simeq -18.5$ mag), photometrically selected LBGs from \hst\ \citep[e.g.,][]{bou10,lab10}. In this work, we have confirmed the existence of such extreme slopes in luminous and spectroscopically confirmed LAEs. We will discuss their nature in the next section.

\section{Discussion} \label{sec:disc}

Generally speaking, higher-redshift galaxies are bluer. For example, Figure \ref{fig:comBeta} shows that $z\ge6$ galaxies have $\beta \le -2$ on average, while low-redshift galaxies typically have $\beta > -1.5$. For LBGs at $z>4$, some studies also found that the same redshift-dependence is still valid, and lower-luminosity galaxies tend to be bluer \citep[e.g.,][]{bou12}. These results are controversial \citep[e.g.,][]{dun13}. In addition, the existence of a population of $\beta \simeq -3$ LBGs at $z\ge6$ are also in debate, as we introduced in Section 1. A slope of $\beta \le -2.7$ usually means a very young stellar population with low metallicity and no dust content. But UV colors cannot be arbitrarily blue for a given initial mass function (IMF). Theoretically, it is difficult to explain $\beta \le -2.7$ using standard stellar population models when nebular emission is included \citep[e.g.,][]{rai10}. For example, the bluest slopes are roughly $\beta \simeq -2.6$ from the BEAGLE code \citep{che16,wil18}. The implications of extreme slopes have been discussed in the literature \citep[e.g.,][]{bou10,wil11}. 

\subsection{SED modeling}

\begin{figure}[t]
\includegraphics[width=8.5cm]{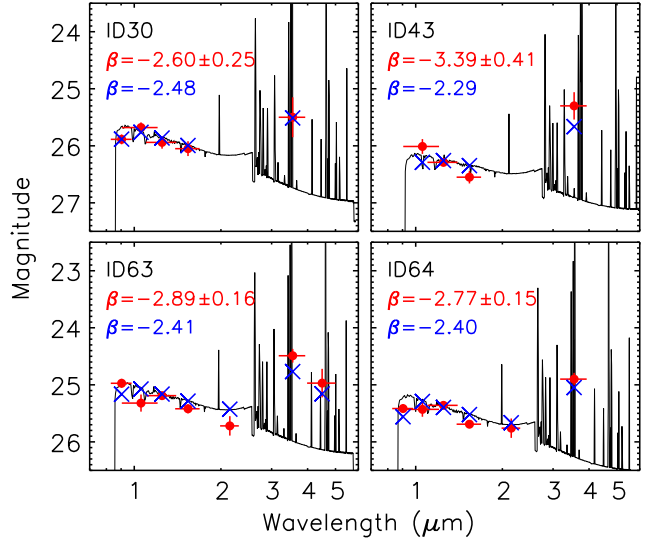}
\caption{SED modeling of four galaxies using GALEV \citep{kot09}. The red points with error bars are the observed photometric data points. The horizontal errors indicate the wavelength ranges of the filters. The black spectra represent the best models. The large blue crosses represent the photometric points predicted by the models. The \lya\ contributions have been removed from the observed photometry (the first data points). The best-fitted UV slopes (in blue) are significantly redder than the observed values (in red) for ID43, ID63, and ID64.
\label{fig:SEDmod1}}
\end{figure}

In order to understand the nature of our galaxies with $\beta < -2.6$, we model their SEDs and explore the possible ranges of a few stellar population parameters. We include the IRAC data from \citet{jiang16}. All galaxies except ID07 were detected in at least one IRAC band. We fit the broadband SEDs using the GALEV evolutionary synthesis models \citep{kot09}. We adopt a Kroupa (similar to Chabrier) IMF with a mass range of 0.1--100 $M_{\sun}$, and assume a constant star formation rate (SFR). Metallicity is fixed to be 0.2 $Z_{\sun}$, a typical value in high-redshift galaxies \citep{fin12}. We use the Calzetti reddening law \citep{cal00} and include metallicity-dependent gaseous or nebular emission. The inclusion of nebular emission is important, because our wavelength range covers very strong emission lines (e.g., \ha, \hb, or \oiii) that affect the broadband SEDs. Note that our spectroscopic redshifts remove one critical free parameter `redshift' and ensure that strong emission lines are covered by proper filters. We mainly constrain three quantities: dust reddening {\it E}(B--V), SED age, and stellar mass. The best fits of four example galaxies are illustrated in Figure \ref{fig:SEDmod1}. 

Our results show that the dust reddening is consistent with zero and the typical age is $\sim$4 Myr (the minimum allowed age) for all five galaxies with $\beta \le -2.6$ and IRAC detections. Two galaxies ID28 and ID30 with $\beta \approx -2.6$ are well fitted. However, it becomes more difficult to fit the bluer galaxies with this model. This is demonstrated in Figure \ref{fig:SEDmod1}, where the best-fitted UV colors are significantly redder than the observed values. The best-fitted slopes $\beta$ are calculated using the same method as we did for the observed data, i.e., we first compute broadband photometry using the best-fitted spectra, and then estimate $\beta$ from the broadband photometry (blue crosses in the figure). From the model spectra, we note that the continuum slopes at $\ge$2 $\mu$m (rest-frame $\ge$3000 \AA\ for $z\simeq6$) are slightly flatter than those at shorter wavelengths. In addition, only ID63 and ID64 (both at $z\simeq6$) have $K_{\rm s}$-band data. For consistency, we do not use the $K_{\rm s}$ band when we calculate slopes $\beta$ from the model spectra. Therefore, the rest-frame wavelength range used here is very similar to the originally defined range of 1250--2600 \AA\ \citep{cal94}. The observed slopes of ID63 and ID64 nearly remain the same without the $K_{\rm s}$-band data points.

Previous studies have shown that standard models cannot produce $\beta < -2.7$. We adopt a top-heavy IMF by increasing $M_{\rm up}$ to 120 $M_{\sun}$. But this barely changes our results, because very massive stars are efficient ionizing sources that produce large amounts of nebular emission, which results in redder colors. The presence of strong nebular line emission is strongly suggested by our IRAC detections (Figure \ref{fig:SEDmod1}). 

\subsection{Further exploration}

\begin{figure}[t]
\includegraphics[width=8.5cm]{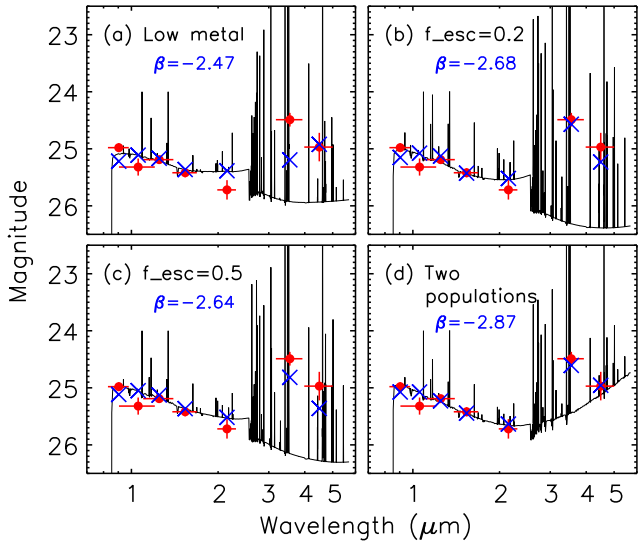}
\caption{SED modeling of ID63 using CIGALE \citep{noll09,serra11,boq19}. The meanings of the symbols are the same as Figure \ref{fig:SEDmod1}. For purpose of clarity, emission lines above 24 mag at $\lambda<2$ $\mu$m in the model spectra have been cut. The combination of a young population and an old population can potentially explain $\beta\simeq-2.9$, shown in (d). The details of the four panels are explained in Section 4.2.
\label{fig:SEDmod2}}
\end{figure}

We use the CIGALE code \citep{noll09,serra11,boq19} to explore further possibilities, including lower metallicities, non-zero escape fractions ($f_{\rm esc}$) of Lyman continuum flux, and AGN contributions ($f_{\rm AGN}$). We modify the code to remove \lya\ emission lines since \lya\ emission is not a free parameter for our galaxies. We adopt a Chabrier IMF with a mass range of 0.1--100 $M_{\sun}$ and assume a constant SFR.  We allow age to vary between 1 and 50 Myr, in steps of 1 Myr. We focus on ID63 that has more observed data points than the others. The results are shown in Figure \ref{fig:SEDmod2}. The best-fitted dust extinction is all (nearly) zero, and ages are 1--4 Myr. In panel (a), we try two low metallicities 0.02 and 0.005 $Z_{\sun}$, and find that the IRAC bands are poorly fitted. ID63 is much brighter in IRAC~1 than it is in IRAC~2. Many such galaxies have been found, and it is believed that the IRAC~1 flux is boosted by strong \oiii\ line emission at this redshift. At a very low metallicity, however, the O/H ratio is reduced, which cannot produce the (IRAC~1 -- IRAC~2) color seen in ID63.

\citet{bou10} have used the \citet{sch03} models to explain $\beta \simeq -3$. These models involve some extreme populations such as Population III and those with extremely low metallicities ($<0.001\, Z_{\sun}$) and IMFs with $M_{\rm up}=500\, M_{\sun}$. When a metallicity is as low as $\le 0.001\, Z_{\sun}$, $\beta$ can be as blue as --3.0 over a small age range of $10\sim50$ Myr. \citet{bou10} argue that this may explain some of the galaxies with $\beta\simeq-3$. However, it does not likely apply to ID63, since such a low metallicity cannot explain its (IRAC~1 -- IRAC~2) color. It is worth pointing out that we have used the same metallicity for the continuum and nebular emission in our models. The gas-phase nebular abundances could be much higher than the stellar metallicity \citep[e.g.,][]{ste16}. We have not tried to use difference abundances for the two components. 

In the above models, we have assumed the Lyman continuum escape fraction $f_{\rm esc}=0$, which is consistent with previous observations that the majority of the low-redshift galaxies observed so far have $f_{\rm esc}<5$\%. A much higher $f_{\rm esc}$ ($\ge20$\%) is required to reionize the Universe at $z>6$. Meanwhile, a higher $f_{\rm esc}$ can apparently make a steeper slope, not only because a higher fraction of very blue stellar light directly leaks from the galaxy, but also because less UV radiation produces less nebular emission. We use $f_{\rm esc}=0.2$ with a metallicity of 0.2 $Z_{\sun}$ \citep[e.g.,][]{smith18}, and we show the result in panel (b) of Figure \ref{fig:SEDmod2}. We obtain a better fit with $\beta\simeq-2.7$, which can nearly explain the observed $\beta\simeq-2.89\pm0.15$. We further increase $f_{\rm esc}$ to 0.5, and obtain a  worse fit, shown in panel (c). This is because there are not sufficient ionizing photons that produce strong nebular line emission to match the IRAC flux. In the simplest case, a line intensity is proportional to $n_i \times n_e$, where $n_i$ is the ion (atom) responsible for the line emission and $n_e$ is the electron density. As we mentioned earlier, the IRAC~1 and IRAC~2 flux is mainly boosted by \oiii\ and \ha. An increased $f_{\rm esc}$ or a decreased amount of ionizing photons means smaller $n_i$ and $n_e$ for \ha\ or smaller $n_e$ for \oiii. Therefore, we cannot arbitrarily increase $f_{\rm esc}$. 

We have not considered AGN contributions so far. Our galaxies are spectroscopically confirmed and do not show type 1 AGN features, such as broad line emission. However, the existing X-ray data are not deep enough to fully rule out AGN activities in such faint objects. We add a small AGN contribution $f_{\rm AGN}=0.1$, with an assumption of $f_{\rm esc}=0.2$. The AGN model has a central power source with a surrounding dust torus. The spectrum of the central source is a broken-power law, with slope $\beta=0.2$, --1, and --1.5 at 10 \AA\ $ < \lambda < 300$ \AA, 300 \AA\ $ < \lambda < 1250$ \AA, and 1250 \AA\ $ < \lambda < 20$  $\mu$m, respectively \citep{fritz06}. We take the nominal values for the torus parameters \citep{boq19}. This is a coarse estimate, since we do not have observational constraints on the AGN contribution. The result is comparable to the case without the AGN contribution.

Finally, we try the combination of a young and an old stellar populations. This is motivated by the fact that a young population with a high $f_{\rm esc}$ can produce a very steep slope, but with weak IRAC flux (e.g., in case of ID63). The IRAC flux can potentially be boosted by an old population. Recent studies have shown that two populations may exist at $z\ge6$, including an evolved population and a recent starburst likely induced by galaxy merging \citep[e.g.,][]{has18}. Given the limited number of the data points, we are not able to fully explore the parameter space. Instead, we make a simple case that includes a 1-Myr old, dust-free population with $f_{\rm esc}=0.6$, and a 650-Myr old population with $f_{\rm esc}=0$ and {\it E}(B--V)=0.5. In this case, the young population dominates the wavelength range blueward of the Balmer break, and the contribution of the old population is negligible in this range. Both populations contribute to the IRAC flux redward of the Balmer break. We obtain a reasonable result with $\beta\simeq-2.9$ for ID63, shown in Figure \ref{fig:SEDmod2} (d). Although this is not a real multi-population SED modeling, it provides a hint that the combination of different populations may be a solution to produce extreme slopes. A more comprehensive study is needed to physically explore this possibility. 

For ID43 with $\beta \simeq -3.4\pm0.4$, we are not able to find a reasonable fit to explain its extreme slope. Based on the above models with single populations, the bluest slopes that we can achieve is $\beta \simeq -2.7$ when nebular emission is included. In the two-population model, we can obtain $\beta \simeq -2.9$. Steeper slopes cannot be well explained by these models. On the other hand, the measurement errors are still large. Deeper observations are needed to reduce the errors.

\section{Summary} \label{sec:sum}

We have carried out a study of six very blue LAEs at $z\simeq6$. These galaxies are luminous and spectroscopically confirmed, selected from a large galaxy sample that has previous \hst\ near-IR and {\it Spitzer} IRAC observations. In \hst\ Cycle 25, we obtained new WFC3 images. The combination of the new data and existing data have allowed us to reliably measure the UV-continuum slopes using 3--4 photometric data points. The slopes are in the range of $-3.4<\beta<-2.6$, with a typical error of 0.2--0.4. In particular, four galaxies have $\beta\le-2.8$. Such blue, luminous, spectroscopically confirmed galaxies have not been reported previously. We performed SED modeling using the multi-band data. The two $\beta\approx-2.6$ galaxies can be well fitted using standard stellar populations with very young (a few Myr) ages and zero dust content. It is more difficult to fit bluer galaxies using single stellar populations. The bluest slopes that the models predict are $\beta\simeq-2.7$. We note that our IRAC detections have put strong constraints on $f_{\rm esc}$ and metallicity, such that a larger $f_{\rm esc}$ cannot produce strong nebular line emission as evidenced by the IRAC flux, and an extremely low metallicity cannot produce the observed (IRAC~1 -- IRAC~2) color for ID63. The inclusion of a small AGN contribution does not change our conclusion. Finally, we found that the combination of a young and an old populations can potentially explain the observed SEDs of the $\beta \simeq -2.9$ galaxies. In this case, the young population has zero dust content and a high $f_{\rm esc}=0.6$, which produces a very blue slope and dominates the $\beta$ measurement. The old, dusty population with a strong Balmer break contributes little to $\beta$. Both young and old populations contribute to the IRAC flux. This model is not able to produce UV slopes as blue as $\beta=-3.4\pm0.4$ in the remaining two galaxies. 

In order to fully understand these galaxies, further spectroscopic observations are required. In particular, they are expected to have strong nebular emission lines in the mid-IR. Given the brightness of these galaxies, JWST will be able to directly detect their continuum and line emission in the near future.

\acknowledgments

We acknowledge the support from the National Key R\&D Program of China (2016YFA0400703) and the National Science Foundation of China (11721303, 11890693). The US investigators acknowledge grants HST-GO-15137 from STScI, which is operated by AURA for NASA under contract NAS 5-26555. RAW acknowledges support from NASA JWST Interdisciplinary Scientist grants NAG5-12460, NNX14AN10G and 80NSSC18K0200 from GSFC. We also thank the anonymous referee for helpful suggestions.

\facilities{HST(WFC3)}

\software{SExtractor, DrizzlePac, CIGALE}



\end{document}